# Fomite transmission and disinfection strategies for SARS-CoV-2 and related viruses

Nicolas Castaño[+], Seth Cordts[+], Myra Kurosu Jalil[+], Kevin Zhang[+], Saisneha Koppaka, Alison Bick, Rajorshi Paul, and Sindy KY Tang*

Department of Mechanical Engineering, Stanford University

[+]equal contributions
*sindy@stanford.edu

**ABSTRACT**

Contaminated objects or surfaces, referred to as fomites, play a critical role in the spread of viruses, including SARS-CoV-2, the virus responsible for the COVID-19 pandemic. The long persistence of viruses (hours to days) on surfaces calls for an urgent need for surface disinfection strategies to intercept virus transmission and the spread of the disease. Elucidating the physicochemical processes and surface science underlying the adsorption and transfer of virus between surfaces, as well as their inactivation, are important in understanding how the disease is transmitted, and in developing effective interception strategies. This review aims to summarize the current knowledge and underlying physicochemical processes of virus transmission, in particular via fomites, and common disinfection approaches. Gaps in knowledge and needs for further research are also identified. The review focuses on SARS-CoV-2, but will supplement the discussions with related viruses.

## 1. Introduction

As of May 21, 2020, the COVID-19 pandemic has infected >5M and caused >328K deaths worldwide,[1] significantly more than MERS and SARS combined.[2] In general, the primary routes of respiratory virus transmission are: 1) direct contact between individuals, 2) indirect contact via contaminated objects, also referred to as fomites, 3) airborne transmission via droplets and aerosols.[3]

There is growing consensus that contaminated fomites play a critical role in the spread of viruses,[4] in a wide range of environments including hospitals, schools, offices, restaurants, and nursing homes.[4,5] Surface disinfection is one of the strategies to intercept fomite-based disease transmission. For example, UV irradiation has been applied to disinfect vehicles for public transportation.[6] Mass spray chemical disinfection is also currently used in multiple countries,[7–9] by using robots, unmanned aerial vehicles, and other semi-autonomous or autonomous spray equipment. While these disinfection strategies can have multiple potential benefits, the effectiveness of some methods, especially for the control of COVID-19, is unexamined.

Most reports on COVID-19 fomite transmission and interception measures are focused on empirical results. Understanding the science underpinning the interactions between viruses and surfaces, as well as their inactivation, is equally important for a better understanding of how the disease is transmitted, how to intercept the transmission, and eventually how to devise guidelines to prevent the spread of the disease.

The goals of this paper are to: 1) review the current knowledge and underlying physicochemical processes of virus transmission, in particular via fomites, and common disinfection strategies, and 2) identify gaps in knowledge and needs for further research. The review focuses on SARS-CoV-2, the virus responsible for COVID-19, but will supplement the discussions with related viruses, as much about SARS-CoV-2 is still unknown. We hope that the review can provide value by stimulating research efforts to further our understanding of the transmission of the disease, as well as by facilitating the development and implementation of effective disinfection strategies.

The remainder of this review is organized as follows. Section 2 provides a short summary of the structure of SARS-CoV-2 as relevant to its transmission via fomites and inactivation. Section 3 summarizes the different routes of virus transmission, how viral load is quantified, and efforts in modeling the infection risks via these routes. Section 4 focuses on virus transmission via fomites, starting with a brief discussion of the physicochemical origin of virus adsorption to and transfer between surfaces, followed by empirical findings of the transfer rate and persistence of viruses on different surfaces. Section 5 reviews the inactivation mechanisms of viruses, and current strategies to intercept fomite transmission. Section 6 examines methods of applying chemical disinfectants and their effectiveness.

## 2. SARS-CoV-2 structure, envelope properties and surrogate viruses

Here we focus on the properties of SARS-CoV-2 that are directly relevant to their transmission via fomites and disinfection strategies. The details of its structure, as well as other coronaviruses, can be found in previous work.[10–12]

### 2.1 Basic Structure

SARS-CoV-2 is a betacoronavirus, which is a genus of enveloped viruses with a linear, positive sense, single-stranded RNA genome that encodes for four main structural proteins: envelope (E), membrane (M), spike (S), and nucleocapsid (N).[10] SARS-CoV-2, similar to other enveloped viruses, is composed of two structures: 1) a lipid bilayer envelope that surrounds 2) a nucleocapsid, a protein capsid enclosing the genome strand.[13]

The E, M, and S proteins are embedded in the lipid bilayer envelope. This lipid bilayer is derived from the host cell and is formed during the budding of a nucleocapsid through a cell membrane.[11] The lipid bilayer is susceptible to chemical disruption, for example, by surfactants. Disruption of the lipid envelope could render the virus inactive.[14] In addition to the lipid layer, the M and E proteins could be targets for the inactivation or weakening of SARS-CoV-2 due to their critical roles in viral envelope assembly and replication. While enveloped viruses are more susceptible to inactivation than non-enveloped viruses, they possess the ability to adapt the envelope molecular profile to evade immune systems.[14,15]

### 2.2 Physicochemical properties

While the exact size of SARS-CoV-2 has not been reported, the approximate diameter of the closely related SARS-CoV-1 is 82-94 nm, with spikes that extend ~19 nm out (total diameter of ~120-132 nm).[16] The isoelectric point (pI) of viruses is important in determining their adsorption characteristics. Based on the protein composition, the pI of the M and N proteins of other coronaviruses have been computed theoretically to be ~9.3-10.7.[17–19] The pI of the M and N proteins on SARS-CoV-2 are likely to be within the same range. Although the overall isoelectric points of coronaviruses have not been reported, they are expected to be largely influenced by the isoelectric properties of M and N proteins,[19,20] and can be further approximated by accounting for the dissociation constants of all amino acids of the virus.[21]

### 2.3 Surrogates

SARS-CoV-2 requires biosafety level (BSL) 3 facilities to handle.[22] To facilitate the investigation of its infectivity, transmission, and disinfection, it is useful to identify surrogates with similar structures to SARS-CoV-2 but with reduced risk of human infection. The first class of surrogates involves the use of natural viruses with low infectivity in humans. Table 1 shows these surrogate viruses, host cells, and BSL levels reported thus far.

**Table 1:** SARS-CoV-1/SARS-CoV-2 Surrogate viruses

| Surrogate viruses | Virus Family | Host cells | BSL | References |
|---|---|---|---|---|
| TGEV (transmissible gastroenteritis virus) | Coronaviridae | Swine testicular cells (ST cells)<br>Porcine kidney cells (PK15 cells) | BSL 2 | [28–31] |
| PEDV (Porcine epidemic diarrhoea virus) | Coronaviridae | African green monkey kidney cells (Vero 76 cells)<br>Porcine intestinal epithelial cell line (IPEC-J2 cells) | BSL 2 | [31,32] |
| MHV (mouse hepatitis virus) | Coronaviridae | Mouse epithelial cell line (NCTC)<br>Mouse delayed brain tumor cell (DBT cells) | BSL 2 | [28–30,33] |
| CCV (canine coronavirus) | Coronaviridae | Dog fibroblast cells (A-72 cells) | BSL 2 | [34,35] |
| Phi6 | Cystoviridae | Pseudomonas syringae | BSL 1 | [36–39] |
| MS2 | Leviviridae | Escherichia coli | BSL 1 | [36–40] |
| Qβ | Leviviridae | Escherichia coli | BSL 1 | [39,40] |

A second class of surrogates is pseudotyped viruses. They are derived from parent viruses such as the murine leukemia virus (MLV), human immunodeficiency virus (HIV), and herpes simplex virus (HSV). The genome of the parents are modified for safer use in BSL 2 labs.[23] The synthesis of pseudotyped viruses is highly adaptable and allows for the incorporation of various kinds of envelope glycoproteins.[24,25] For example, SARS-CoV-2 S glycoprotein has been incorporated into a lentiviral pseudotyped virion system to determine the potential drug targets for the virus.[26]

A third class of surrogates involves artificial capsids that emulate the viral architecture. For example, peptide capsids have been constructed using capsid proteins to serve as non-pathogenic viral surrogates.[27] They have been used to study aspects of viral infectivity, applied as antimicrobial agents to disrupt bacterial lipid bilayer membranes, and programmed to carry specific genetic cargo and deliver into the cytoplasm of human cells.

## 3. Virus transmission: routes, measurements, and modeling

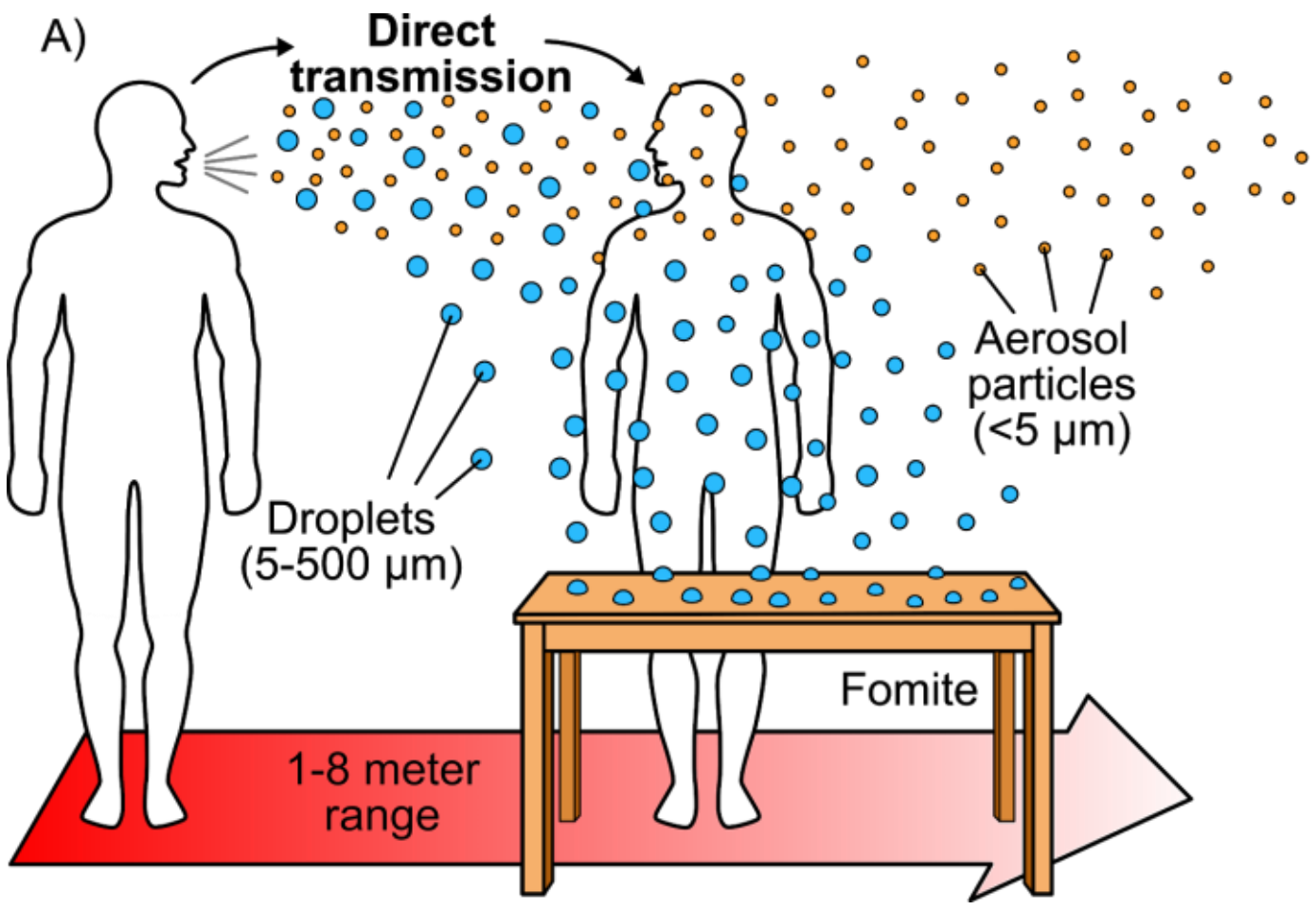

**Figure 1.** A) Respiratory droplet (up to 500 µm) and aerosol (<5 µm) particles produced by an infected host during coughing/sneezing, talking, or exhaling can infect fomites or another individual directly. Droplets settle and adsorb onto fomites while aerosol particles can remain suspended in air for minutes to hours.[41,42] B) Indirect fomite-mediated transmission to a new human host occurs through contact with the fomite and subsequent contact with regions through which a virus can enter the body. Contact times can range from ~1-50 s.[43]

### 3.1 Transmission routes of viruses

Understanding the transmission routes of viruses is crucial to the development of effective control measures. Three primary transmission routes have been found to contribute to the spread of respiratory viruses (e.g., SARS-CoV-1 and -2, measles, HCoV, rhinovirus, and influenza virus) (Figure 1A): 1) direct contact between individuals, 2) indirect contact via contaminated objects (fomites), 3) airborne transmission via droplets and aerosols.[3]

Direct contact involves the transmission of the virus through physical contact between an infected host and a susceptible individual. Direct contact is a potent transmission route since the viral load can be large, and the virus spends a shorter amount of time outside of a host compared with other routes of transmission. For MERS, SARS-CoV-1, and SARS-CoV-2, direct contact is considered a major transmission route.[44,45]

Indirect contact involves the transmission of the virus through a contaminated intermediate object called a fomite. Fomites are inanimate objects or surfaces that can become contaminated by the physical contact with either another infected fomite or skin, or by settling airborne particles. Fomite transmission can occur when an individual touches a contaminated fomite, and then touches their facial membranes (Figure 1B). Numerous studies have implicated fomites as a significant virus transmission route in a range of environments.[4,5] Although the transfer efficiency of SARS-CoV-2 from fomites to other fomites or skin is not well characterized, the transfer efficiency of a number of viruses has been investigated, and will be detailed in Section 4.

Respiratory viruses can also become airborne and spread via particles generated by sneezing, coughing, talking, or exhaling. The particles generated can be classified into droplets or aerosols based on a cut-off diameter of 5 µm.[3] These airborne droplets and aerosols can cause infection through inhalation into the respiratory tract,[42] or by settling onto fomites.[4] Due to their large size, droplets typically fall within 1-2 m of the source within seconds (Figure 2).[42,46–48] Aerosols are smaller and can remain suspended in the air for minutes to hours (Figure 2), depending on the environmental conditions.[41,42] Such prolonged suspension could increase the distance the virus travels from the source, and the number of individuals and fomites exposed to the virus.

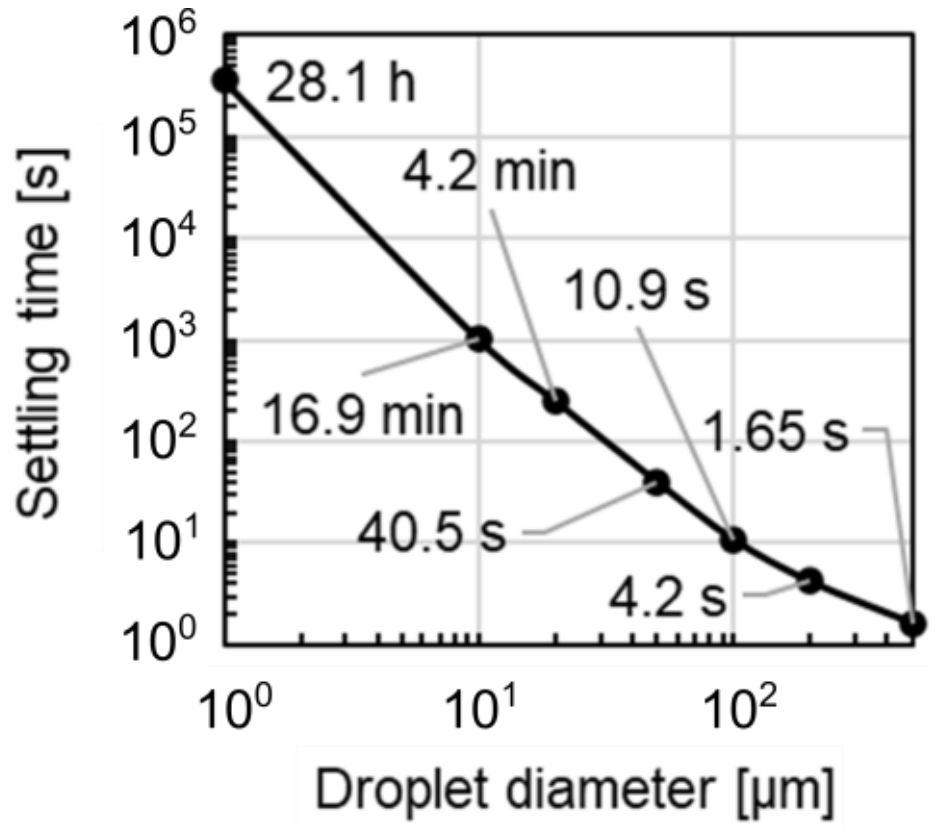

**Figure 2.** Droplet settling time from a height of 3 m was approximated based on its terminal velocity.[49] In this approximation, settling time scales with the second power of droplet diameter and the air around the droplet is assumed to be stagnant.

Detailed experimental and theoretical approaches have estimated that more violent events such as sneezing can deposit droplets and aerosols up to 6 – 8 m away from the source.[47,48] Following the initial respiratory event, nearby air currents (e.g., from ventilation or wind) can re-suspend aerosols and extend the range over which aerosols can travel.[50] Although the degree to which SARS-CoV-2 can be transmitted by aerosol remains unclear, early evidence from a study in Wuhan hospitals reported that the virus was detected in aerosol samples from areas open to the general public, ~$10^1$ – $10^2$ meters from the source.[51] Airborne droplets and aerosols can also deposit onto and contaminate fomites. A study on SARS-CoV-2 infected patients in isolation rooms showed contamination of high-contact surfaces such as doorknobs and bedrails, as well as air outlet fans which indicated virus transfer from aerosols to a surface.[52]

While the transmission routes discussed above are generally accepted as the main transmission routes of respiratory viruses, sewage and dust-borne transmission have also been implicated as possible routes. SARS-CoV-1 and SARS-CoV-2 have both been detected in sewage, suggesting the possibility of transmission via the fecal-oral route, or via aerosolization of sewage caused by flushing.[53,54] Furthermore, dust-borne transmission has been proposed as a possible mechanism in the spread of avian influenza in chickens.[55]

The relative importance of each transmission route in the spread of SARS-CoV-2 and most respiratory viruses remains an open question. A trend that is generally accepted, though, is that the risk of infection increases for persons who are in close proximity to an infected individual for extended periods of time. For example, the probability of transmission of SARS-CoV-2 between family members and close contacts[44] and among passengers and workers on cruise ships[56] were much higher than that among the general population. Additionally, a ferret model showed high transmission efficiencies of SARS-CoV-2 among ferrets living in close quarters, while ferrets separated by a permeable partition were infected less efficiently.[57]

Because of the uncertainty or unavailability of quantitative data, it is difficult to draw conclusions about how each transmission route contributes to the increased risk of infection in close proximity. Furthermore, there is often a confounding effect between transmission routes. For example, persons in sufficiently close proximity for droplet-based transmission are likely exposed to a great intensity of virus-laden aerosols simultaneously.[58] In a model of influenza infection assessing the relative contributions of each transmission route to infection risk within a household, fomite transmission was estimated as a major, if not the dominant, transmission route.[59] Although the relative importance of each transmission route remains poorly understood, there has been a growing consensus that contaminated fomites play a critical role in the spread of viruses.[4]

### 3.2 Measurement of virus infectivity

In order to determine the viability of a virus on surfaces and in aerosols, it is crucial that the methods of collecting virus particles are effective in representing the virus titer from the environmental sample accurately. Additionally, the methods used to analyze the collected samples must have high specificity and sensitivity. In this section, we summarize the current collection and analysis methods, as well as opportunities for improvement.

#### 3.2.1 Aerosol sample collection

The most common methods of collecting aerosolized virus particles use a gelatin filter because it is highly efficient in collecting virus particles without affecting viral infectivity.[60] Gelatin filters can also be dissolved easily for harvesting, culturing, and quantifying live virus particles. In a recent study, the stability of SARS-CoV-1 and SARS-CoV-2 in aerosols was measured by generating an aerosol and feeding the aerosol into a Goldberg 25 drum containing a gelatin filter.[61] Alternative methods for collecting aerosolized viruses have been covered in a previous review.[60] Briefly, these methods use solid impactors (e.g., Andersen sampler, slit sampler, and cyclone sampler), liquid impactors (e.g., all-glass impingers, swirling aerosol collector), filters (e.g., gelatin, cellulose, polycarbonate, PTFE, or cotton), or electrostatic precipitators. The downside of these approaches is the significant time delay (minutes to hours) between particle collection and virus quantification. In an effort to develop real-time methods for virus aerosol collection and detection, microfluidics-based assays have been developed. Briefly, these methods include microfluidic optical immunosensing of latex agglutination,[62,63] aerosol detection by impingement onto a microfluidic droplet detector,[64] and label-free virus capture using carbon nanotubes and detection by Raman spectroscopy.[65] However, these approaches are still in development and are not yet validated for wide-spread use.

#### 3.2.2 Surface sample collection

No study has evaluated the efficiency of virus collection from an exhaustive list of surfaces. In a recent study examining commonly used collection methods, the most effective method for recovering MS2 bacteriophages from nonporous fomites used polyester-tipped swabs pre-wetted in either one-quarter-strength Ringer's solution or saline solution.[66] This method recovered a median fraction of 0.40 for infective MS2, and 0.07 for MS2 RNA from stainless steel and PVC surfaces.

#### 3.2.3 Quantification of infectious virus and RNA concentration

The two prevailing methods for detecting viral RNA and viable virus particles are reverse transcriptase polymer chain reactions (RT-PCR) and plaque assays, respectively.

RT-PCR assays are used to determine both the presence of viral RNA and the number of copies of viral RNA. While various forms of RT-PCR exist (e.g. rRT-PCR, qRT-PCR, and LAMP-PCR), they all follow a general principle of amplifying specific viral RNA for detection and quantification. RT-PCR assays are well characterized, straight-forward to perform, and do not require cell culture. Their limitation is the inability to determine virus infectivity.[67]

Plaque assays involve culturing cells that are susceptible to virus infection in a titration of the collected virus samples, monitoring the cytopathic effects, and counting plaque forming units (PFU). Several recent preprint publications have used Vero E6 cells to quantify the presence of infective SARS-CoV-1 and Sars-CoV-2.[61,68–71] While plaque assays are the most popular method for determining infectivity, they have several limitations, including: 1) the propensity to human error, 2) the long duration of the assay (possibly exceeding a week) due to the time required for observable cytopathic effects and/or PFUs, 3) the lack of a reliable host cell model for some viruses, and 4) the absence of plaque formation in some viruses.[67] To address some of these limitations, alternative approaches have been developed. One alternative is the 50% tissue culture infectious dose ($TCID_{50}$), an endpoint dilution assay that quantifies the infectious titer required to produce cytopathic effects in 50% of a tissue culture. However, the $TCID_{50}$ assay is also susceptible to some of the limitations of the plaque assay (e.g., long durations before cytopathic effects are detectable).[72]

To improve the detection limit of plaque assays, integrated cell culture followed by quantitative PCR (ICC-qPCR) has been used. The number of infective virus particles is enhanced for detection by first culturing the virus particles with host cells. The virus is then extracted for RT-PCR to quantify the amount of viral RNA. Samples that had infective virus particles produced a larger end-point value during RT-PCR than samples that did not have viable virus. ICC-qPCR has been used for many virus strains.[67,73–76]

An alternative method to detect viable viruses without cell culture is to pretreat the collected samples with proteinase K and RNase before performing RT-PCR. If the virus envelope is damaged (i.e., the virus is non-infective), proteinase K and RNase will break down the capsid and any RNA that is not in a stable viral envelope. However, this method still requires optimization and broader validation.[67,77]

Other detection methods exist but have not been fully developed or validated for coronavirus. For example, ELISA and immunofluorescence assays that take 15-30 minutes have been developed for influenza virus detection. While these tests have high specificity (98.2% from bivariate random-effects regression), they have modest sensitivity (62.3%) only.[78]

As can be seen, most current methods take a few hours to measure virus RNA concentration and days to measure virus infectivity. Additionally, most methods require instrumentation and/or cell culture that may be inaccessible. There is a critical need for the rapid detection (< hours) of viable infective virus not only from patient samples, but also from aerosols, droplets, and fomites to better understand the infection risk via these transmission routes.

### 3.3 Modeling of infection risks

Mathematical models of infection risk can be useful to estimate the relative importance of transmission routes and to evaluate the effectiveness of preventative measures. Here we discuss several existing models of an individual's risk of infection from their immediate environment. A more detailed review can be found elsewhere.[79] Large-scale models at the community level are beyond the scope of this review, but can be found in prior work.[80,81]

Models of infection risk often consist of two tasks: estimating the viral load, or dose, received, and estimating the infection risk based on the dose received. Infection risk models can be classified as deterministic, where an individual is infected if the dose exceeds a critical value, or stochastic, where an individual's probability of infection is a function of the dose received. Typically, stochastic models are more biologically relevant.[79]

To estimate the dose received, various strategies exist to model the transmission routes of the virus from the surroundings to the individual. In the case of aerosol transmission, a Poisson distribution is often used to describe the distribution of virus particles in the air.[79] This distribution can be used to estimate the dose an individual receives through aerosol inhalation.[82] For the Wells-Riley model (discussed later), the airborne “dose” is formulated in terms of a hypothetical unit called a quantum of infection.[83,84] The Wells-Riley “dose” includes parameters such as the quanta generation rate, room ventilation rate, and exposure time to describe aerosol transmission and can be further modified to account for complexities such as uneven mixing.[79]

In the case of fomite transmission, some models estimate the surface concentration through deposition from an airborne source,[85] while others directly prescribe a distribution on a surface based on experimental measurements.[86] The rate of virus transferred between two surfaces is often formulated as the product of contact frequency, transfer efficiency, surface concentration, and contact area in Eq. 1,

$$\dot{N}_{surface\ 2} = f_{1,2} \times \varepsilon_{1,2} \times C_{surface\ 1} \times A_{contact\ 1,2} \qquad \text{(Eq. 1)}$$

where $\dot{N}_{surface\ 2}$ is the rate of virus transferred to surface 2, $f_{1,2}$is the frequency of contact between surfaces 1 and 2, $\varepsilon_{1,2}$ is the transfer efficiency from surface 1 to 2, $C_{surface\ 1}$ is the virus concentration on surface 1, and $A_{contact\ 1,2}$ is the contact area between surface 1 and 2.

The virus can be transferred serially between surfaces before finally transferring to an individual’s facial membranes, giving the final dose received through fomite transmission.[86] More complex models integrate multiple transmission routes and phenomena together, such as a Markov chain model of fomite and aerosol transmission that include a gradual loss of virus viability.[87]

To estimate the infection risk based on the dose received, two popular models have emerged: the Wells-Riley model and the dose-response model. The Wells-Riley curve is an exponential curve and is based on a hypothetical “dose” unit called a quantum of infection as described above.[83,84] A basic form of the Wells-Riley curve is shown in Eq. 2,

$$P = 1 - exp\left(-\frac{Iqpt}{Q}\right) \qquad \text{(Eq. 2)}$$

where $P$ is the probability of infection, $I$ is the number of infectors, $q$ is the quanta generation rate, $p$ is the pulmonary ventilation rate of the susceptible individual, $t$ is the exposure time interval, and $Q$ is the room ventilation rate with clean air.[79] While the Wells-Riley model is convenient to use, its formulation based on the quantum of infection limits its application to aerosol transmission only. Still, it remains a useful model and has been applied to various cases including SARS-CoV-1.[88]

The dose-response model was adapted for respiratory viruses from models of toxicity.[82] The dose input is a physical quantity of the virus, and can be extended to multiple situations including aerosol transmission,[82] fomite transmission,[89] and the efficacy of surface disinfection strategies.[86] A basic form of an exponential dose-response curve is shown in Eq. 3,

$$P = 1 - exp\left(-\frac{IGp\beta t}{Q}\right) \qquad \text{(Eq. 3)}$$

where $P$ is the probability of infection, $I$ is the number of infectors, $G$ is the number of airborne pathogens released per infector per unit time, $p$ is the pulmonary ventilation rate of the susceptible individual, $\beta$ is the deposition fraction of pathogens in the alveolar region, $t$ is the exposure time interval, and $Q$ is the room ventilation rate with clean air. Extensions and alternative forms of the dose-response curve can be found elsewhere.[79,90]

Some studies have begun to apply these models of individual infection risk to a larger system such as a household. For example, interactions between multiple healthy individuals, infected individuals, and objects in a household can be modelled and the dose-response model is then applied to each individual.[59,91] Ultimately, the choice of the model depends on the application, and models of infection risks have emphasized the effectiveness of promising interventions including fomite disinfection.[85,86]

## 4. Virus transmission via fomites

The ability of a virus to transfer between and persist on different surfaces, including skin, plays a crucial role in the overall infectivity of a virus by means of fomite transmission. Understanding the adsorption and transfer rates between skin and fomites is critical for modeling the spread of viruses.[5,92] Furthermore, understanding virus persistence on different surfaces under different environmental factors can inform decision making for disinfection protocols. In this section, we will review the factors affecting virus adsorption, transfer, and persistence on different surfaces, and then discuss surfaces that are at high risk of contamination.

### 4.1 Physicochemical origin of virus adsorption and transfer

The adsorption of virus on fomites and their subsequent transfer to other surfaces is a multi-factor problem that depends on the properties of the virus, the fomite, and the environment.

The physical description of virus adsorption borrows from formulations of colloid adsorption, treating virus particles as soft colloidal spheres and using Gibbs free energy to model the interactions between virus particles and the adsorbing surface. Like colloid adsorption onto surfaces, virus adsorption onto fomites is primarily driven by electrostatic, hydrophobic, and van der Waals interactions (Figure 3). The relative contribution of these interactions is modulated by environment pH and ionic strength.[4,92,93]

Classical models of virus adsorption adopt the Derjaguin–Landau–Verwey–Overbeek (DLVO) theory for colloid adsorption onto surfaces. It accounts for electrostatic and van der Waals interactions between viruses and surfaces.[94–96] However, the extended-DLVO (XDLVO) model, which considers hydrophobic interactions, was found to agree more with experimental observations of virus adsorption.[97–99] XDLVO is expressed in terms of Gibbs free energy of interaction, shown in Eq. 4,

$$\Delta G_{total} = \Delta G_{dl} + \Delta G_{vdW} + \Delta G_{hyd} - T\Delta S_0 \qquad \text{(Eq. 4)}$$

where electrostatic or double-layer (dl), hydrophobic (hyd), and van der Waals (vdW) contributions are summed. Entropy changes ($\Delta S_0$) are usually ignored. A negative $\Delta G_{total}$ favors adsorption.[95,96] Detailed formulations for each component of the total free energy for a spherical virus particle adsorbing onto a flat plate can be found in prior work.[98]

Electrostatic forces drive long-range adsorption dynamics dictated by the radius of the virus's electrical double-layer (Debye length) and the charge of the absorbing surface.[94,100] All viruses, including SARS-CoV-2, express unique protein markers on their surfaces. These markers consist of weakly acidic or basic polypeptides and amino acids ionizing residues that give viruses characteristic isoelectric points (pI) (also see section 2.2).[94,101] The net charge of a virus is thus determined by the pH of its environment.[94] The net charge on a virus causes the formation of an electrical double-layer that extends from the Stern layer, the first layer of immobile charges attached to the surface of the virus particles, and across the Gouy diffuse layer, the region of charge imbalance that results in an electrical potential.[102–104] In addition to pH, the ionic strength of the surrounding medium is another important parameter affecting electrostatic interaction. At high ionic concentrations (>100 mM $NaNO_3$)[100], electrostatic screening stunts the zeta potential at the charge slipping plane and weakens the effects of surface charge for both attractive and repulsive interactions.

In the absence of electrostatic interaction, if adsorption occurs, it is typically attributed to hydrophobic interactions.[100,102] The hydrophobic effect causes an attractive force between a virus and adsorbing surface. The effect is due to electron-donor and -acceptor, i.e., Lewis acid-base, interfacial interactions. Under hydrophobic interactions, there is a tendency of apolar species such as molecular chains or particles to aggregate,[96] providing an energetically favorable mechanism of adsorption due to the minimization of interfacial area between the virus and the adsorbing material.[40] In the absence of electrostatic interactions, hydrophobic effects dominate because they are apolar by nature. The energy of hydrophobic interactions largely depends on the prevalence of hydrophobic groups on a virus particle's surface. Greater virus hydrophobicity has been shown to correlate with higher rates of adsorption regardless of ionic strength.[100,105,106] The presence of chaotropic (i.e., $SCN^-$, $Cl_3CCOO^-$) or anti-chaotropic (i.e., $NO_3^-$, $SO_4^{2-}$, $F^-$) agents can promote or hinder, respectively, hydrophobic adsorption.[94] Hydrophobic interactions are considered to have a short-range effect compared with electrostatics.

Van der Waals forces are considered to be of secondary importance. Their relative contribution, as with electrostatic and hydrophobic interaction, is a function of virus and environmental properties. For example, van der Waals forces may play a significant role in the adsorption of viruses that carry a neutral charge in their environment.[107] Furthermore, materials known to generate large van der Waals potentials are also more likely to adsorb viruses.[94] The contribution of van der Waals forces to adsorption can be quantified by Lifshitz theory, which predicts that materials with higher dielectric susceptibility produce higher van der Waals potentials. By this reasoning, metals have better adsorbing effectiveness than most organic substances. In general, the effectiveness of materials to absorb viruses follows: Metals > Sulfides > Transition Metal Oxides > $SiO_2$ > Organics. This theory suggests that high ionic strength or a fluid pH equal to virus pI is necessary for adsorption to most organics. Under these conditions, the Debye length is shortened and viruses are able to get sufficiently near to organic surfaces to absorb by van der Waals interactions.[94,95,108]

There exist some gaps in the comprehensive understanding of the physicochemical mechanisms in virus adsorption onto fomites. While XDLVO theory on virus adsorption can begin to explain the observations of many virus strains, including SARS-CoV-2, readily adsorbing to a variety of non-porous surfaces (e.g., steel, glass, plastic),[4,61,68,109,110] the observed virus adsorption onto porous fomite surfaces (e.g., cardboard, cloth) is not well described by XDLVO. Some studies have indicated the need to account for steric effects and surface roughness.[40,106] Other studies emphasized the pitfalls of modeling viruses as soft colloids with homogeneous charge distributions. Unlike a soft colloid or even a virus-like particle (VLP) engineered with viral structural proteins, the pI of true viruses depends on the complex physicochemical structure of the outer surface and the genetic material packed within the capsid.[100,107,111,112]

The state of understanding in physicochemical mechanisms of virus adsorption in aqueous environments is fairly advanced, but there is still significant room for research in elucidating the mechanisms of dry contact transfer. Although a number of studies have quantified the rates of transfer between dry surfaces, including skin (also see Section 4.2), the precise physicochemical basis for virus transfer in dry conditions has been unaddressed. The tendency of a virus to transfer between fomites is likely determined by differences in adsorption energies between the two surfaces. In the case of porous materials, lower rates of transfer are likely due to viruses entrapped in their matrix due to increased surface area for attachment.[113,114]

To our knowledge, no work has examined the physicochemical interactions, adsorption, and transfer kinetics of SARS-CoV-2 on different surfaces especially in dry conditions.

### 4.2 Transfer efficiency of viruses between skin and other surfaces

Despite a lack of data on the transfer efficiency of coronaviruses, numerous studies have examined the transfer efficiency for other viruses. An overall mean transfer efficiency of 23% ± 22% was found between fingerpads (either washed or unwashed prior to inoculation with a virus) and glass for three types of non-enveloped bacteriophages (MS2, φX174, and fr). The efficiency was calculated by measuring the viral PFU of the surface before and after contact. In this study, prior handwashing was found to reduce the transfer efficiencies only slightly. The reduction due to washing was greater in fingerpad-to-glass transmission than glass-to-fingerpad. This result is likely because of changes in the skin moisture or pH due to handwashing before inoculation with a virus.[115] A similar transfer efficiency was found for MS2 from fingertips to glass and to acrylic (~20%), but this value increased to 79.5% in humid conditions.[114] Transfer efficiency of PSD-1 phage from hand to mouth was found to be 33.9%, representing a skin to skin pathway.[116]

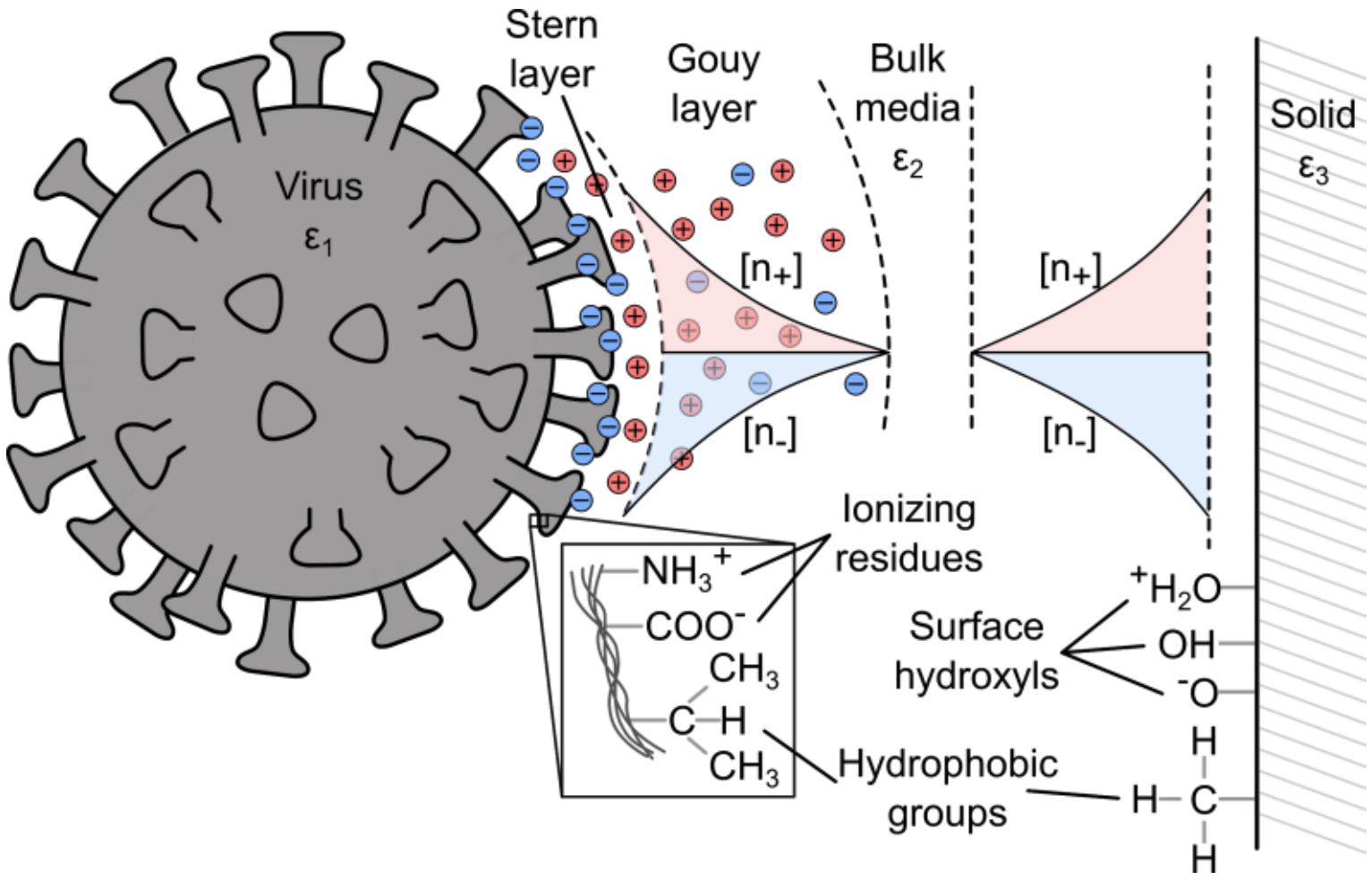

**Figure 3.** Diagram summarizing components contributing to XDLVO-based interactions between a virus and a surface. Ionizing residues on viral amino acids interact with surface hydroxyls groups on an adsorbing surface. The Gouy layer forms from a local imbalance in charge concentration. These long-range electrostatic forces are attractive or repulsive based on the charges on the virus and the surface. Apolar hydrophobic groups on the virus and surface exhibit shorter-range interactions. The complex dielectric susceptibility (ε) mismatch of the virus, media, and solid surface drives van der Waals interactions. Adapted from Gerba.[94]

### 4.3 Persistence of coronaviruses on different surfaces

Studies have shown that viruses adsorbed on surfaces can maintain high rates of survival and infection potential. Exactly how long viruses retain their viability on a surface is highly variable and dependent on: 1) surface porosity, 2) environmental factors, and 3) virus envelope characteristics.[113,117]

First, nonporous surfaces, compared with porous surfaces, are more effective in receiving and transferring viruses, and are typically better at preserving virus viability because they do not draw moisture away from adsorbed viruses.[118] However, if a porous material is inoculated, it is capable of harboring most strains of viruses (especially at low temperatures e.g., 4℃), and can remain contagious despite the lower rates of transfer to skin.[113] SARS-CoV-2 has demonstrated an ability to contaminate a wide range of porous and nonporous fomites. Table 2 shows the persistence of SARS-CoV-2 and other coronaviruses on various surfaces. To our knowledge, no studies have evaluated the persistence of SARS-CoV-1 or SARS-CoV-2 on skin. However, parainfluenza was shown to be 5% viable within 10 minutes, while rhinovirus was shown to be 37.8% and 16% viable on skin after 1 and 3 hours, respectively. The percent of viability was determined by comparing PFU before inoculation and at various times points.[119] We note that no work has explicitly investigated the physicochemical reasons why some surfaces support longer virus persistence. As viruses can be inactivated by desiccation,[29] improved persistence is likely due to the ability of a surface to maintain a moist microenvironment.

Second, environmental variables such as temperature, humidity, and resident microfauna can influence virus adsorption and viability. In general, increased temperature and moderate humidity levels have adverse effects on the persistence and viability of coronaviruses and other viruses.[32] In a study on the viability of dried SARS-CoV-1 on smooth plastic surfaces, the virus was found to be viable for over 5 days at 22-25°C with 40-50% relative humidity (RH). However, virus viability

**Table 2:** Summary of the persistence of human coronaviruses along with porcine and murine coronaviruses, TGEV and MHV, respectively, which are commonly used as surrogates for human coronavirus. High levels of discrepancies in viability between similar virus-surface combinations could be attributed to experimental differences in environmental conditions and inoculum titer.

| Virus | Type/ strain | Quantification method | Inoculum titer | Surface type | Relative humidity | Temperature | Viability period | Ref. |
|---|---|---|---|---|---|---|---|---|
| SARS-CoV | Type 2 | Vero-E6 cell plaque assay and RT-PCR | 5 µL of $10^{7.8}$ $TCID_{50}$ /mL | Cloth | 65% | 22°C | ≤ 2 days | 68 |
| | | | | Steel | | | ≤ 7 days | |
| | | | | Glass | | | ≤ 4 days | |
| | | | | Plastic | | | ≤ 7 days | |
| | | | | Wood | | | ≤ 2 days | |
| | | | | Bank note | | | ≤ 4 days | |
| | | | | Paper, tissue paper | | | ≤ 3 hr | |
| | | | | Surgical mask | | | ≤ 7 days | |
| | Type 2/ nCoV-WA1-202 | Vero-E6 cell plaque assay | 50 µL of $10^5$ $TCID_{50}$/mL | Steel | 40% | 21-23°C | ≤ 72 hr | 123 |
| | | | | Copper | | | ≤ 4 hr | |
| | | | | Plastic | | | ≤ 72 hr | |
| | | | | Cardboard | | | ≤ 24 hr | |
| | Type 1/ Tor2 | Vero-E6 cell plaque assay | 50 µL of $10^5$ $TCID_{50}$/mL | Steel | 40% | 21-23°C | ≤ 72 hr | 123 |
| | | | | Copper | | | ≤ 8 hr | |
| | | | | Plastic | | | ≤ 48 hr | |
| | | | | Cardboard | | | ≤ 8 hr | |
| MERS-CoV | Isolate HCoV-EMC/ 2012 | Vero-E6 cell plaque assay | 5 µL $10^6$ $TCID_{50}$/mL | Steel, Plastic | 40% | 20°C | ≤ 48 hr | 124 |
| | | | | | 80% | 30°C | ≤ 8 hr | |
| | | | | | 30% | 30°C | ≤ 24 hr | |

(Table 2 continued)

| Virus | Type/ strain | Quantification method | Inoculum titer | Surface type | Relative humidity | Temperature | Viability period | Ref. |
|---|---|---|---|---|---|---|---|---|
| HCoV | 229E | L132 cell plaque assay | 10 µL of 5.5 x $10^5$ $TCID_{10}$/mL | Aluminum, sterile sponge, latex glove | 55-75% | 22°C | ≤ 3 hr | 125 |
| | | MRC-5 cell plaque assay | 10 µL $10^3$ PFU/$cm^2$ | Glass, PVC, teflon, steel | 30-40% | 21°C | ≤ 5 days | 126 |
| | | | | Rubber silicon | | | ≤ 3 days | |
| | | | | Copper nickel (> 90% copper) | | | < 30 min | |
| | OC43 | HRT-18 cell plaque assay | 10 µL of 5.5 x $10^5$ $TCID_{50}$/mL | Aluminum, sterile sponge, latex glove | 55-75% | 22°C | < 1 hr | 125 |
| TGEV | Not specified | Swine testicular cell plaque assay | 10 µL of $10^4$-$10^5$ MPN/$cm^2$ (MPN is the most probable number of virus particles) | Steel | 20-80% | 4°C | > 28 days | 29 |
| | | | | | 20-80% | 20°C | 3-28 days | |
| | | | | | 20-80% | 40°C | 4-96 hr | |
| MHV | Not specified | Delayed brain tumor cell plaque assay | 10 µL of $10^4$-$10^5$ MPN/$cm^2$ | Steel | 20-80% | 4°C | > 28 days | 29 |
| | | | | | 20-80% | 20°C | 4-28 days | |
| | | | | | 20-80% | 40°C | 4-96 hr | |

decreased significantly (>3 $log_{10}$ reduction) at 38°C with >95% RH. [120] In another study using Phi6 as a surrogate, the virus survived best at high (>85%) and low (<60%) RHs. They also found that RH is a more significant factor in virus survivability than absolute humidity (AH).[36] In addition to temperature and humidity, the presence of other microbes can also influence the survival of viruses. Though the presence of microbes can reduce the rate of desiccation of the viral particles enhancing their persistence and viability, microbial proteases and fungal enzymes can be harmful to their existence.[121,122]

Third, viral persistence on fomites also depends on the type and the strain of the virus. In general, non-enveloped enteric viruses (e.g., adenovirus, rotavirus) can persist on fomites longer than enveloped viruses (e.g., coronaviruses).[4] The lack of a lipid membrane in non-enveloped viruses make them less susceptible to inactivation than enveloped viruses, where the disintegration of the lipid envelope (e.g., by common disinfectants; see details in section 5) causes the loss of the viral envelope proteins involved in virus adsorption and cell penetration thereby rendering them inactive.[14] Additionally, non-enveloped viruses are less susceptible to desiccation than their enveloped counterparts because of their lack of lipid membrane envelopes. These characteristics make them easier to spread and persist on surfaces over long periods of time compared with enveloped viruses.[113]

### 4.4 Surfaces at high risk of virus contamination

In principle, all surfaces or objects can be considered potential fomites and are at risk of contamination.[4,5] In practice, knowledge of which objects are at high-risk of contamination could guide the design of optimal disinfection strategies. For a given object, the risk of contamination can depend on the interaction between the virus and the material, the frequency at which the object is contacted, the object's distance from an infected individual, and the environmental conditions.

First, the combination of virus composition and surface properties can influence the likelihood of contamination (see details in Section 4.1). Second, objects that are frequently handled or are in high contact with individuals are at higher risk of contamination. In a hospital setting, contamination has been detected on numerous high-contact surfaces, including door handles, bed rails, tables, call/control panels, other near-patient surfaces, office equipment, and even sterile packaging.[85,127] A study of the isolation rooms of SARS-CoV-2 infected patients in Singapore showed contamination of a similar list of high-contact surfaces.[52] While the floor of the isolation room and the shoes worn by individuals entering and exiting the room tested positive for SARS-CoV-2, the floor immediately outside tested negative, suggesting contamination by footwear is low.

Third, an object's proximity to an infected individual affects its risk of contamination. An object can be contaminated from a distance due to deposition of droplets or aerosols onto its surface. The risk of contamination by droplets or aerosols decreases when the object is further away from infected individuals, as viral shedding by coughing, sneezing, or exhaling can potentially deposit droplets and aerosols onto fomites as far as 6 - 8 m away.[47,48] In the aforementioned Singapore study, all air samples taken from the isolation room tested negative while the air outlet fans tested positive, suggesting that SARS-CoV-2 is not detectably aerosolized in these conditions but is still able to transfer from air to a potential fomite.[52] A study in Wuhan hospitals found that the highest concentrations of SARS-CoV-2 in the air were, surprisingly, not in patient rooms but in toilet facilities.[51]. Even aerosol generation from personal protective equipment (PPE) removal can create fomites. Doffing PPE has the potential to aerosolize the virus and transfer it to other PPE in changing rooms.[51]

Fourth, the environmental conditions can affect an object's risk of contamination. Air currents could potentially determine the flight path of droplets and aerosols, as proposed in a case study of a Guangzhou restaurant where the SARS-CoV-2 infection pattern aligned with the air conditioning currents.[128] The amount of foot traffic and the degree of connectivity between rooms could also affect where high SARS-CoV2 concentrations may be found.[129] We note a limitation to many of these studies is the use of RT-PCR to identify viral RNA. The presence of viral RNA is not indicative of viability, and viral culture is needed to determine infective virus [66].

The above factors can be used to help identify and predict surfaces at high-risk of contamination. To further quantify the role of these surfaces as fomites, surface viral concentrations need to be measured, and contact frequencies can be derived from observational studies.[130] Such quantifications can be used as input parameters in modeling infection risk and designing optimal disinfection strategies. For example, a model of disinfecting strategies for methicillin-resistant *Staphylococcus aureus* (MRSA) predicted once-daily whole room cleaning to be less efficient than frequent targeting of high contact surfaces in preventing indirect contact transmission.[85]

## 5. Current strategies to intercept fomite transmission route

Strategies to intercept fomite transmission revolve around inactivating the virus, improving personal hygiene, or using PPE. Here, we discuss different mechanisms of virus inactivation on surfaces and hands, focusing on strategies that have been shown to inactivate SARS-CoV-2 and other enveloped viruses. We will not discuss PPE as it has been discussed elsewhere.[131–140]

### 5.1 Reactivity of viral structures with disinfecting agents

In order for a virus to be infective, it must fuse with a host cell, insert its genome into the cell, and replicate.[141] These processes require an enveloped virus to have an intact envelope and nucleocapsid. To inactivate a virus, at least one of these components needs to be disrupted.[67]

It is important to understand the mechanisms of virus inactivation based on virus composition, structure and function in order to: 1) understand the efficacy of disinfectants on viruses, 2) predict the response of a new strain of virus to a disinfectant, 3) identify common sites on proteins, envelopes or genomes that are vulnerable to disinfectant treatment that are shared by many viruses, and 4) enhance the design of antiviral agents and therapies that target specific viral components.[142]

One method of predicting virus inactivation mechanisms is a composition-based method. It takes into account the reaction rate constants between disinfectants and specific nucleotides and amino acids found in viral structures (e.g., proteins, nucleic acids, and envelope lipids) (Table 3). The relative contribution of a viral structure to inactivation can be predicted by summing the relative abundance of each nucleotide or amino acid multiplied by the respective rate constants for a given disinfectant.[21,142] A limitation of the composition-based method is that it does not account for the complex interactions between adjacent monomers.[142] To our knowledge, no model exists that accounts for viral structure as well as composition.

### 5.2 Surface disinfection strategies

This section summarizes current disinfection strategies and their effectiveness for SARS-CoV-2 and related viruses (Figure 4).

#### 5.2.1 Ultraviolet (UV) and solar irradiation

UV irradiation is a widely-used method of surface disinfection. Here, we discuss the inactivation of SARS-CoV-1 by UVC and solar irradiation.

UVC irradiation (100 - 280 nm, typically 254 nm is used) damages nucleic acid bases in genetic material, and to a lesser extent, proteins in virus capsids.[67,141] UVC irradiation induces dimerization of adjacent uracil bases in RNA, forming pyrimidine dimers that disrupt the RNA structure, which inhibits the viral replication process and inactivates the virus.[67,153] Exposure of SARS-CoV-1 to a UVC light source (254 nm, 4016 µW/cm$^2$) held 3 cm above the virus resulted in a ~4.5 $\log_{10}$ $TCID_{50}$/mL reduction in virus titer within 6 minutes. After this point, virus inactivation plateaued, until viral activity became undetectable at 15 minutes. Exposure to other UV wavelengths (e.g., UVA) was found to be insufficient to inactivate the virus.[154]

UVC irradiation as a disinfection strategy poses a few challenges. The time required to inactivate SARS-CoV-1 using UVC (254 nm, 4016 µW/cm$^2$), ~6 minutes, is significantly longer than the time required using chemical disinfectants (30 s to 1 min).[110,154] This time to inactivate only applies to regions of an object directly exposed to UVC irradiation. Disinfectant effectiveness reduces significantly in shadowed regions. Additionally, UVC radiation may pose health risks, including skin cancer and ocular damage to exposed individuals.[155] Nonetheless, UVC-based disinfection can be valuable for use in applications where the irradiation can be shielded from humans, and have been used in, for example, empty buses and other vehicles.[6]

The inactivation of viruses by solar irradiation has also been studied, especially in the context of the disinfection of water. The range of UV wavelengths in sunlight that reach the surface of the earth is between 290 and 400 nm, as UVC is typically completely blocked by the atmosphere.[153,156] The antiviral properties of sunlight primarily come from UVB light, which can also form pyrimidine dimers, but these mechanisms are not as well studied as the mechanisms of UVC-based disinfection.[153] Additionally, the solar spectrum, especially in the UV wavelengths, can vary significantly depending on environmental factors, the time of day, and the season. Such factors can lead to large variations in the efficiency of virus inactivation by sunlight.[153]

#### 5.2.2 Chemical disinfectants

A wide variety of chemical disinfectants are currently available to combat the spread of SARS-CoV-2 (Table 4).[110] The effectiveness varies depending on the virus inactivation mechanism. In general, there are 3 modes of inactivation by disinfectants: 1) disruption of the lipid layer of the envelope (e.g., ethanol and detergents),[67,143] 2) modification of important protein sites on the envelope or capsid (e.g., chlorine and glutaraldehyde),[67,141,142,149,157] and 3) reaction with the nucleotides and amino acids in the genetic material, leading to the degradation of the nucleic acids (e.g., chlorine).[67,141]

Chemical disinfectants are typically evaluated with suspension and carrier tests. Suspension tests combine a known titer of a virus in solution with a disinfectant and evaluate virus titer after a period of time that depends on the disinfectant manufacturer's directions of use.[158] However, suspension tests are considered less challenging for the disinfectant under scrutiny,[159] and may not reflect the practical usage of disinfectants to clean contaminated surfaces. Quantitative carrier tests are performed by allowing an aliquot of virus solution to dry on a surface before applying the disinfectant. This test is conducted under conditions that are more relevant to practical use of disinfectants, and is, therefore, a more appropriate measure of disinfectant effectiveness.[159]

Chemical disinfectants have been evaluated for their ability to inactivate various types of coronavirus.[68,110] For SARS-CoV-2, 1% and 2% household bleach, 70% ethanol, 7.5% povidone-iodine, 0.05% chloroxylenol, 0.05% chlorhexidine, and 0.1% benzalkonium chloride have been found to reduce an initial viral load of ~7.8 $\log_{10}$ $TCID_{50}$/mL to undetectable levels at room temperature within 5 minutes in suspension tests.[68] For other coronaviruses (e.g., SARS-CoV-1, MERS-CoV, and MHV), 78%, 80%, 85% and 95% ethanol;[160–162] 75% 2-propanol;[161] a mixture of 45% 2-propanol and 30% 1-propanol;[162] 0.21% sodium hypochlorite;[163] and 1%, 4% and 7.5% povidone iodine[164] were found to reduce viral activity by at least 4 $\log_{10}$ within 30 seconds in suspension tests.

Carrier tests performed on stainless steel disks showed >3 $\log_{10}$ reduction within 1 minute for HCoV-229E, MHV, and TGEV exposed to 70% ethanol[30] and for HCoV-229E exposed to 0.1-0.5% sodium hypochlorite and 2% glutaraldehyde.[165] In another study, hydrogen peroxide vapor inactivated TGEV in a carrier test by a reduction of 4.9-5.3 $\log_{10}$, but it took 2-3 hours to do so.[166] Prior carrier tests have been performed primarily on stainless steel, and to our knowledge, no

**Table 3.** List of disinfectants, their reactivity with monomers of virus structures: nucleotides and amino acids. Table adapted from Wigginton and Kohn.[142]

| | Nucleotides and amino acids | Disinfectant type | | |
|---|---|---|---|---|
| | | UVC (254 nm) | Free chlorine | Ozone |
| Reactivity with nucleotides Second order rate constant, k [$M^{-1}s^{-1}$] reported for chlorine and ozone<br><br>Molar attenuation coefficient, ε [$M^{-1}cm^{-1}$] reported for UVC. | Adenine | 1.2 x $10^4$ | 6.4 | 200 |
| | Cytosine | 3.5 x $10^3$ | 66 | 1.4 x $10^3$ |
| | Guanine | 1.0 x $10^4$ | 2.1 x $10^4$ | 5.0 x $10^4$ |
| | Uracil | 7.8 x $10^3$ | 5.5 x $10^3$ | 650 |
| | Thymine | 6.3 x $10^3$ | 4.3 x $10^3$ | 1.6 x $10^4$ |
| Reactivity with amino acids Second order rate constant, k [$M^{-1}s^{-1}$] reported for chlorine and ozone.<br><br>Molar attenuation coefficient, ε [$M^{-1}cm^{-1}$] reported for UVC. | Cysteine | -- | 3.0 x $10^7$ | ~1 x $10^9$ |
| | Histidine | -- | 1.0 x $10^5$ | ~4 x $10^5$ |
| | Lysine | -- | 5.0 x $10^3$ | -- |
| | Methionine | -- | 3.8 x $10^7$ | ~6 x $10^6$ |
| | Phenylalanine | 140 | -- | -- |
| | Tryptophan | 2.8 x $10^3$ | 1.1 x $10^4$ | ~1 x $10^7$ |
| | Tyrosine | 340 | 44 | ~4 x $10^6$ |
| | Backbone N | -- | ≤ 10 | -- |
| | α-amino | -- | 1.0 x $10^5$ | -- |

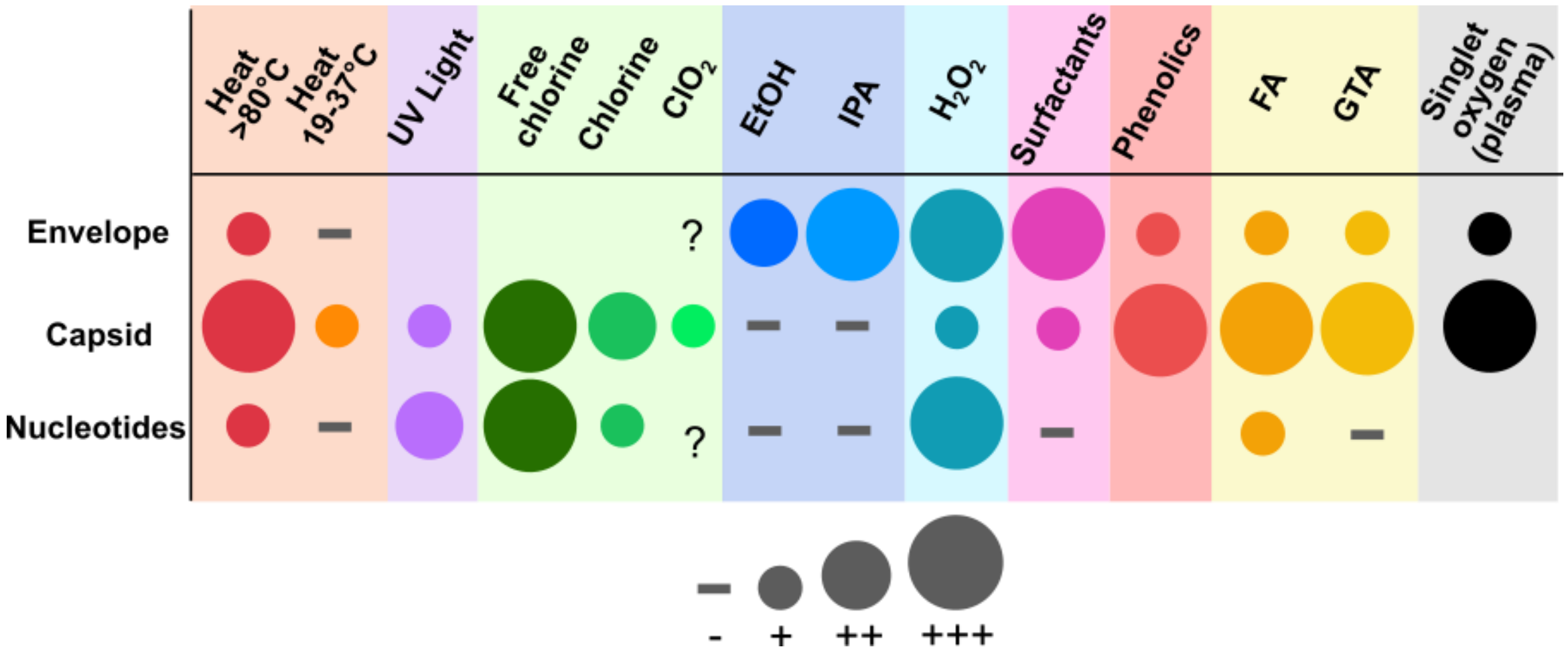

**Figure 4.** Viral structures targeted by different disinfectants. Symbol abbreviations: +, light damage; ++, moderate damage; +++, severe damage; −, no damage; ?, uncertain/debated. Chemical abbreviations: $ClO_2$, Chlorine dioxide; EtOH, Ethanol; IPA, Isopropanol; $H_2O_2$, Hydrogen peroxide; FA, formaldehyde; GTA, glutaraldehyde. Expanded on and adapted from Zhang et al.[67] References: Heat[141,143,144], UV Light[141,144–147], Chlorines[141,148], EtOH and IPA, $H_2O_2$[149], Surfactants[14], Phenolics[149], FA and GTA[150,151], Singlet oxygen[141,146,152].

**Table 4:** Advantages, disadvantages and hazards of disinfectants that have been shown to inactivate SARS-CoV-2 and similar coronaviruses. It is important to ensure that the choice of disinfectant is safe and compatible with the substrate it is applied to.
*NFPA rating specified as [Health, Flammability, Instability].

<table>
<tr><th>Disinfecting agent</th><th>Hazards</th><th colspan="2">Advantages</th><th colspan="2">Disadvantages</th></tr>
<tr><td>UVC Light<br>(245 nm)</td><td>Adverse health effects from irradiation[155]</td><td colspan="2">Inactivates SARS-CoV-1.[154]<br>UV light may be applicable to surfaces sensitive to heat or chemicals.</td><td colspan="2">UV irradiation is less efficient at low temperatures (e.g., <10°C ) and is not suitable for all environments.[167]<br>Incompatible with photosensitive materials.</td></tr>
<tr><td>1-2% Sodium hypochlorite (bleach)</td><td>[3, 0, 1]*<br>12.5 wt%[168]</td><td>0.21% sodium hypochlorite inactivates MHV in 30 sec. [163]</td><td rowspan="6">Inactivates SARS-CoV-2 within 5 min by reduction of ~7.8 $log_{10}$.[68]<br>Inexpensive and readily available.[169]</td><td></td><td rowspan="6">Inactivation effectiveness depends on virus structure.[67]<br>May be incompatible with some metals and may stain substrates.</td></tr>
<tr><td>70% Ethanol</td><td>[2, 3, 0]*<br>100 wt%[170]</td><td>>80% ethanol inactivates SARS-CoV-1 within 30 sec. [162]<br>Leaves no chemical residue.</td><td>May evaporate faster than required contact time for inactivation.</td></tr>
<tr><td>7.5% Povidone-iodine</td><td>[2, 1, 0]*<br>100 wt%[171]</td><td>Inactivates MERS-CoV in 15 sec.[164]</td><td></td></tr>
<tr><td>0.05% Chloroxylenol</td><td>[1, 1, 0]*<br>3 wt%[172]</td><td></td><td></td></tr>
<tr><td>0.05% Chlorhexidine</td><td>[1, 0, 0]*<br>2 wt%[173]</td><td></td><td></td></tr>
<tr><td>0.1% Benzalkonium Chloride</td><td>[3, 1, 0]*<br>50 wt%[174]</td><td></td><td></td></tr>
<tr><td>Heat<br>(>70ºC)</td><td>Igniting combustibles</td><td colspan="2">Inactivates SARS-CoV-2.[68]<br>Sources of heat (ovens and autoclaves) readily available.</td><td colspan="2">Not applicable to heat-sensitive surfaces.</td></tr>
</table>

carrier tests have been conducted for SARS-CoV-2 on any material. Results using stainless steel carriers may not reflect disinfectant effectiveness on other fomites with different surface properties (e.g., surface chemistry, wettability, porosity, roughness). The dependence of disinfectant effectiveness on surface properties remains an open question.

### 5.2.3 Plasma disinfection

Plasma is an ionized gas made up of charged and uncharged particles (i.e., ions and electrons, and molecules and atoms, respectively), reactive species, and UV photons.[175,176] Plasma can be thermal or non-thermal depending on whether electrons are at the same or higher temperature as heavy particles. Cold atmospheric plasma (CAP) is a low-temperature, non-thermal plasma that is produced by a variety of methods using gases such as helium, argon, nitrogen, heliox and/or air. Two common methods for producing CAP are dielectric barrier discharge and atmospheric pressure plasma jet.[175]

CAP has been considered for disinfection applications in dentistry and oncology and in food processing.[175,176] The antimicrobial properties of CAP are attributed to reactive oxygen and nitrogen species generated in the non-thermal plasma.[176–178] The detailed inactivation mechanisms are still under investigation. However, it is believed that the reactive species damage genetic material and proteins.[179] In one study, singlet oxygen in plasma was implicated in the inactivation of bacteriophages through multiple mechanisms involving reactions with amino acids and DNA nucleotide oxidation and crosslinking, but the primary mechanism was thought to be singlet oxygen-induced crosslinking of capsid proteins.[179]

CAP has been shown to inactivate potato virus Y in water, in a study aimed to decontaminate water supply systems. CAP was found to inactivate potato virus Y in 1 minute, compared with 15 minutes using 25 mg/L of hydrogen peroxide.[176] Plasma-activated water (PAW) is another form of plasma-based disinfection. Water is treated with non-thermal plasma to produce PAW, which is more acidic and contains more reactive oxygen and nitrogen species than regular water[180] rendering it antimicrobial. PAW has been proposed as an antiviral agent. For example, in a study testing PAW for its potential application in producing an inactivated vaccine for Newcastle disease (ND), the enveloped virus responsible for avian ND was found to be fully inactivated in PAW.[178] PAW has also been proposed for its potential application in the disinfection of fresh produce.[180] While promising in these studies, the effectiveness of CAP or PAW on inactivating coronaviruses remains to be demonstrated.

#### 5.2.4 Heat Treatment

Heat treatment is a well-known method for disinfecting surfaces. At temperatures exceeding ~80°C, viral capsid proteins are denatured and RNA is damaged.[67] SARS-CoV-2 has been shown to become inactivated within 5 minutes at 70°C, with a reduction from an initial concentration of ~6.8 $\log_{10}$ $TCID_{50}$/mL to undetectable levels.[68] Sufficiently high temperatures should be used. Moderately high temperatures (19-37°C) only cause minor damage to the protein capsid, and fail to inactivate some viruses.[67]

Autoclave is a common method of sterilizing equipment using heat treatment in a laboratory or clinical environment. Autoclaves produce steam at high temperatures (~132°C) in a pressurized chamber. At this temperature, most microbes, including viruses, are inactivated. The surface being sterilized is exposed to the high temperature and pressure environment for a varying amount of time, depending on the material and size of the object. Liquids are usually sterilized for 30-60 minutes,[181] while objects made of glass and plastics require ~30 minutes of sterilization.[182] In one experiment, avian coronavirus and avian pneumovirus carried by cotton swabs were inactivated after heat treatment using an autoclave for 20 minutes. In the same study, heating the same viruses in a microwave oven for 5 seconds was also found to be sufficient for inactivation.[183]

### 5.3 Self-disinfecting materials and surfaces

Engineering self-disinfecting surfaces is an emerging avenue of research for preventing infection transmission by fomites. While certain materials like copper and silver have long been known to possess intrinsic antimicrobial properties, various types of surface modification and functionalization can also give rise to antimicrobial properties against bacteria and viruses.[127,184] Only a limited number of works have focused on virus-specific self-disinfection.[185] This section highlights some of these studies. Readers are referred to a recent review for details.[186]

Copper and silver alloys are known viricidal agents that inactivate viruses through multiple modes of action. The primary mechanism involves direct interaction between metal ions and microbial proteins, or indirect interaction through the formation of radicals that are damaging to DNA and lipid membranes.[187,188] Copper has been shown to retain its effectiveness across a range of humidities and temperatures, while silver had drastically reduced antimicrobial effectiveness at low humidities (~20% RH).[189]

Pure copper and alloys with 79-89% copper were found to be most effective in inactivating viruses. Abrasion and removal of the outer oxide layer caused a slight decrease in effectiveness. In one study, $5 \times 10^5$ PFU/cm$^2$ of non-enveloped murine norovirus was inactivated in under 2 hours at room temperature.[190] Inactivation of norovirus by copper was found to be up to 10x faster in dry conditions compared to wet, but the mechanisms underlying such differences were unclear.[191] In another study, copper yielded a near 4-log reduction in enveloped influenza A virus particle count after 6 hours.[192] Table 2 includes the effectiveness of copper on some coronaviruses in lowering viability periods. Some studies examining the clinical effectiveness of copper surfaces have shown notable improvement towards infection control benchmarks with 94% less bacteria when compared with control plastic surfaces on ICU beds.[193]

Photocatalytic action has been shown to be highly effective in inactivating microbes by damaging DNA and lipid membranes via the photocatalyzed formation of hydroxyl radicals in the presence of photoactive oxides.[194,195] Numerous enveloped and non-enveloped viruses have been shown to become inactivated by photocatalytic disinfection.[194] Titanium dioxide ($TiO_2$) is a popular photocatalytic material due to its long lifetime, effectiveness over a wide range of microbes, and environmental friendliness. $TiO_2$ has the potential to provide antiviral protection to a range of materials. For example, cotton fabrics have been impregnated with $TiO_2$ via magnesium nanoparticle carriers.[196] $TiO_2$ impregnated into resin, fiberglass, and PVC have also been used to coat various surfaces in hospitals, schools, and other public places.[197]

Despite the potentials that self-disinfecting surfaces present, widespread adoption of self-disinfecting surfaces, especially in hospitals, has been limited by three main obstacles. The first is a lack of clinical trials showing their efficacy in practice. Second, the costs associated with upgrading or retrofitting equipment discourages hospitals from taking initiatives to introduce self-disinfecting surfaces, though the savings from decreasing nosocomial infections could offset this cost. Third, characterization of effectiveness over repeated cycles has not yet been quantified.[198]

### 5.4 Hand hygiene

Frequent handwashing can lower the incidence of transfer from fomites to facial membranes via contact.[199] Considering the frequency that adults touch their faces (23 times per hour) and the risk of infection that is associated with face touching, handwashing is a critically important personal hygiene habit.[43] In a hospital setting, the WHO recommends 5 critical moments for healthcare workers to wash hands: 1) before contact with a patient, 2) before a cleaning procedure, 3) after exposure to bodily fluids, 4) after contact with a patient, and 5) after contact with fomites surrounding patients.[200]

Although virus transfer to hands is only mildly reduced after recent handwashing,[115], handwashing is effective in reducing the spread of a virus from hands.[199,201] However, handwashing is only as effective as the frequency, the effectiveness of the antiseptic, and thoroughness.[201] The CDC recommends washing for a minimum of 20 seconds.[202] This recommendation was based on a few empirical studies,[203–205] including one that investigated handwashing practices such as wash time (15 s vs. 30 s) and effect of soiled hands on infectivity reduction.[206]

To evaluate the effectiveness of a handwashing, a fingerpad method is typically used.[207,208] Here a virus is inoculated on pre-cleaned fingerpads, allowed to dry, then subjected to exposure to an antiseptic by static contact with the fingerpad.[209,210] The ASTM specifies that an effective handwashing antiseptic must yield a minimum reduction of 4 $\log_{10}$ (99.99%) in virus titer from the initial inoculation titer. However, this standard does not specify a minimum contact time between the fingerpad and the antiseptic.[211] Another potential drawback of these standard tests is that they may not be representative of *in vivo* handwashing behavior of healthcare workers or the general public.[208]

Viruses present a unique challenge for handwashing in that their structure and ability to survive on skin may evade inactivation by handwashing methods customized for bacterial disinfection.[199] Alcohol and isopropanol-based antiseptics (60-80% ethanol) are the most effective non-hazardous antiseptic, especially against enveloped viruses.[200] Other WHO-recommended antiviral antiseptics (from the most to the least effective) are iodophors (0.5-10%), chlorhexidine (0.5-4%), and chloroxylenol (0.5-4%), all of which are less effective than alcohol.[212] In regard to hand sanitizers, SARS-CoV-1 has been confirmed to be the most susceptible to ethanol and isopropanol using suspension tests with 1 part virus at $10^7$ $TCID_{50}$/mL, 1 part media, and 8 parts by volume of an ethanol- or isopropanol-based WHO-recommended antiseptic formulation. A >4 $\log_{10}$ SARS-CoV-1 reduction was achieved in 30 seconds using ethanol and isopropanol formulations at 80% and 75% concentrations, respectively, and using dilutions as low as 40%.[161,200]

## 6. Methods of applying chemical disinfectants

Chemical disinfection remains one of the most commonly used methods for virus disinfection. The effectiveness of chemical disinfection depends on the disinfectant contact time, the surface properties of the fomite, and other environmental factors. As such, how the chemical disinfectant is applied has a significant impact on the disinfection effectiveness. This section discusses two common methods of disinfectant application: wiping, and spraying.

### 6.1 Wiping

For chemical disinfectants to work properly, they must be directly applied to the target surface. The most straightforward and conventional method of applying a chemical disinfectant to a surface is to use a manual wipe. Manual wiping utilizes both physical removal of viruses (which may not kill the viruses), and chemical activity of the disinfectant.[213] The chemical disinfectant can be applied immediately before wiping, or the wipe can be packaged and pre-wetted with the disinfectant. The wipe process typically takes seconds to complete.[213] Despite its convenience, manual wiping is limited by human error and cross-contamination between surfaces.[214,215]

If the proper protocols are followed, manual wiping can effectively disinfect surfaces contaminated with norovirus,[216,217] adenovirus [217] polyomavirus,[217] and numerous bacteria including *Staphylococcus aureus* and *Clostridium difficile*.[218] However, the effectiveness of manual wiping depends on multiple factors including the type of wipe, the type of disinfectant, the target pathogen, the wiping technique (e.g., area covered, pressure applied), and the ratio of disinfectant volume to target surface area.[213] Insight into these factors, such as the effectiveness of microfiber wipes,[219] the number of wipe passes over a surface,[220] and adsorption of disinfectants to wipe material,[221] could serve to optimize wipe protocols.

The key limitations of manual wiping arise from human error in the wiping process, and cross-contamination of pathogens. Multiple studies reported that only ~40% of near-patient surfaces in hospitals were cleaned according to policy.[214] If wipes are re-used between surfaces, there is a risk of transferring pathogens between surfaces.[215] These limitations could potentially have serious consequences especially in a hospital setting, and highlight the need for an automated and effective disinfection strategy.

### 6.2 Spraying

While spray disinfectants are commonly used to disinfect surfaces,[222] their effectiveness has not been well characterized.[223] To our knowledge, no comprehensive model has characterized spray disinfection efficiency taking into account aerosol physics, virus heterogeneity, and surface characteristics. Nevertheless, given that disinfection effectiveness depends on disinfectant contact time (and thus disinfectant volume, if the evaporation of disinfectant is fast), the spray characteristics (e.g., spray droplet size and density) and disinfectant droplet deposition on surfaces are critical factors that must be considered in any attempt to evaluate the effectiveness of spray disinfection (Figure 5).

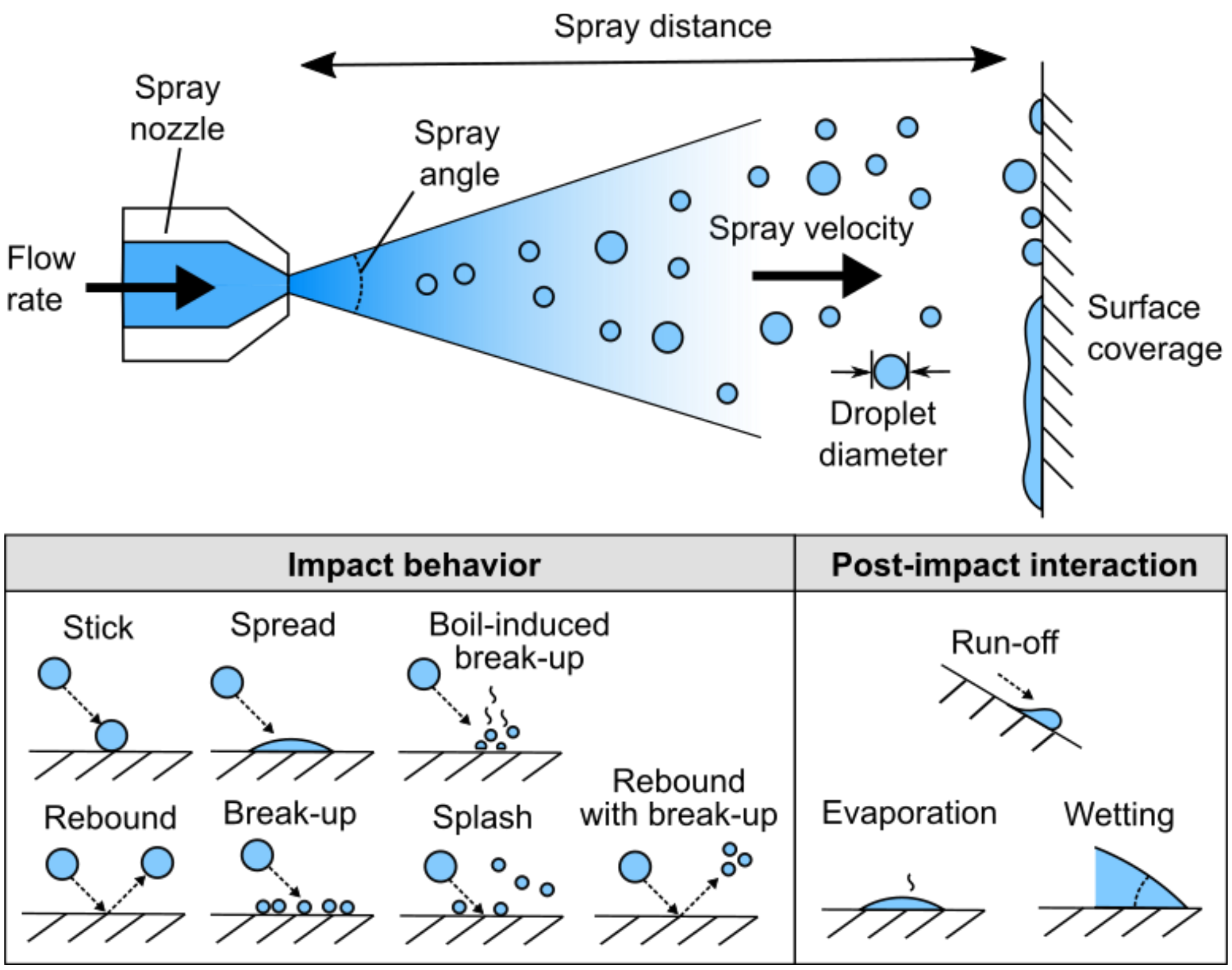

**Figure 5.** Parameters defining spray dynamics. The sprayer/nozzle design affects the spray angle, spray velocity and average droplet diameter. The spray droplets interact with the surface in several manners as shown in the figure. These impact behaviors and post-impact behaviors are dependent on the droplet properties, surface properties, droplet velocity, and temperature. Droplet-surface impact behavior diagrams adapted from Nasr et al.[223]

Atomization is a general term referring to the disintegration of a liquid stream into droplets.[224] Table 5 summarizes a few common commercial atomizers. Atomizers that have been used for applying surface disinfectants (and pesticides) include electrostatic atomizers, hydraulic atomizers, pressure atomizers, spill return atomizers, and ultrasonic atomizers. In particular, spill return atomizers, being able to produce fine sprays, have been shown to be applicable for disinfecting healthcare surfaces.[223] For classroom, healthcare, and general disinfection purposes, handheld electrostatic sprayers also exist with adjustable spray parameters.[225,226] In the agriculture industry, electrostatic spray systems have been mounted on unmanned aerial vehicles for spraying pesticides on crops.[227,228] Drop size and drop-size distribution are critical parameters determining spray disinfection efficiency.

Table 6 summarizes several considerations in the choice of droplet size for spray disinfection.

**Table 5:** Various commercial atomizers and their operating characteristics.

| Atomizer | Typical nozzle diam. | Typical characteristic atomization parameter | Typical droplet size range [μm] | Typical flow rate | Typical spray distance | Ref. |
|---|---|---|---|---|---|---|
| Electrostatic Atomization | 0.75-1.5 mm (fan spray)<br><br>10-20 mm (cone spray) | 1-10 kV (droplet charge) | 25-150 | 0.01-5 L/min | 10-20 ft; 3 ft (for agricultural crop spraying) | [49,227,229–231] |
| Hydraulic Atomization | Varies* | 20-1000 psi | 10 – 5000 | 0.0006 - 1.8 L/min | 4-18 m | [232] |
| Pressure (air) Atomization | Varies* | 100-400 kPa | ≥500 | 0.5-3 L/min | 6-15 in** | [233–235] |
| Rotary Atomizers | 0.75 -710 mm | 10,000-20,000 rpm | 30-50 | 50-500 mL/min | 4-18 in | [236–238] |
| Spill Return Atomizers | 0.3 mm | 9 MPa (supply pressure) | 10-25 | 0.245 L/min | 2.0-2.5 m | [223] |
| Twin-Fluid Atomizers (Internal) | 0.7 - 15 mm | 0.14-0.28 MPa | 100-300 | 1-4 $m^3$/hr | 6-15 in** | [239,240] |
| Ultrasonic Atomizers | 2.5 mm | 20 kHz/ 40 kHz (vibration frequency) | 50-600 | 30 mL/min - 16L/hr | 1-40 cm | [241–243] |

* Parameters vary significantly between manufacturers; no typical values available.
**No typical measurement available. Spray coverage properties given for 6-15” from nozzle in datasheets.

**Table 6:** Pros and cons of small vs. large droplet sizes in spray disinfection.

| Droplet Size | Pros | Cons |
|---|---|---|
| Small (<50 μm) | • Covers a larger surface area for a given volume<br>• Reduced likelihood of coalescence and runoff | • Susceptible to drift<br>• Evaporates faster than large droplets |
| Large (>50 μm) | • Less susceptible to drift and decreased likelihood of undesirable environmental pollution. Typically, droplets with size 50-100 μm are used in practical spray applications such as treating foliage.[49]<br>• Evaporates slower than small drops | • Covers a smaller surface area for a given volume<br>• Increased likelihood of drop coalescence and runoff |

## 7. Conclusions

The COVID-19 pandemic has revealed major gaps in our scientific knowledge, not only in the biology of how the virus infects humans, but also the role of physicochemical processes and surface science in the transmission and inactivation of the virus. Box 1 lists some of the open questions we have identified. Addressing these questions will allow us to devise more effective strategies to combat the spread of the disease. For example, quantitative models predicting the locations of high-risk areas within a building and high-risk objects within those areas can inform the prioritization for disinfection. The identification of surfaces with high contamination risk also presents an opportunity for self-cleaning communal surfaces such as water faucets or door handles. A better understanding of disinfectant effectiveness on different surfaces and their potential side effects allows one to choose the optimal disinfection strategy for specific applications. While our review is by no means exhaustive, we hope that it can provide the basis for researchers in the physical sciences interested in COVID-19 to take on some of the open research challenges, so that as a community we can be better prepared for the next pandemic.

**Box 1: Open questions**

**Fomite transmission of viruses**

- What is the infectivity of the fomite transmission route compared with other routes?
- How can infective viruses be detected in real-time?
- What is the adhesion and transfer efficiency of viruses between human skins and different surfaces?
- How do the surface properties of the virus and surfaces influence the adhesion, transfer, and persistence characteristics?
- How can we better predict the locations and objects that are at high risk of virus contamination?

**Surface disinfection**

- How can one better predict the rate of virus inactivation based on its structure and composition?
- How do the surface properties (e.g., roughness, porosity, wettability, presence of impurities) alter disinfectant effectiveness?
- What is the optimal disinfection strategy to maximize disinfection effectiveness but minimize side effects, including health hazards, pollution, and damage to surfaces?
- What innovations are needed for self-disinfecting surface technologies to be adopted broadly?

## Acknowledgement

We acknowledge support from the National Science Foundation (award #2030390).